\documentclass{aa}
\usepackage{graphicx}
\usepackage{epsfig,times}
\usepackage{color}
\usepackage{amsmath,amssymb}
 \newcommand{\msun}{{\rm M}_{\odot}}
 \newcommand{\rl}{{\rm R_{LC}}}

\begin{document}
\title{Cannonballs in the context of Gamma Ray Bursts } 
\subtitle{Formation sites ?}
\titlerunning{Cannonballs formation sites}
\author{Jan E. Staff\inst{1}, Christian Fendt\inst{2} \and Rachid Ouyed\inst{1,3}}
\institute{Department of Physics and Astronomy,
University of Calgary, 2500 University Drive NW, Calgary, Alberta, T2N 1N4, Canada
\and  Max-Planck-Institut f{\"u}r Astronomie,
K{\"o}nigstuhl 17, D-69117 Heidelberg, Germany
\and Canadian Institute for Theoretical Astrophysics, 60 St. George Street, Toronto,
 Ontario, Canada}
\offprints{jstaff@capca.ucalgary.ca}
\authorrunning{Staff, Fendt, Ouyed}

\abstract{
We investigate possible formation sites of the cannonballs (in the
gamma ray bursts context) by calculating their
physical parameters, such as density, magnetic field and temperature
close to the origin.
Our results favor scenarios where the cannonballs form as
instabilities (knots) within magnetized jets from hyperaccreting disks. 
These instabilities would most likely set in beyond the 
light cylinder where flow velocity with  Lorentz factors
as high as 2000 can be achieved. Our findings challenge the cannonball model of gamma ray bursts if these
indeed form inside core-collapse supernovae (SNe) as suggested
in the literature; 
unless hyperaccreting disks and the corresponding jets are common occurrences 
in core-collapse SNe. 
}
\maketitle

%---------------------------------------------------------------------------------
\section{Introduction}
It has been argued in the literature that, as an alternative to the
fireball scenario (e.g. Piran \cite{piran99}, and references therein),
the so-called cannonball (CB) model provides a good fit to the
observed GRB flux and temporal variations (\cite{dar04}). For example,
 to explain GRBs, CBs must be created in supernova explosions and accelerated
to  high Lorentz factors, $\Gamma_{\rm CB} \sim 1000$.
However, the origin of these highly relativistic ``balls'' of matter 
has not yet been investigated and is the subject of much debate and controversy.
 In order to  shed some light on the still open questions
 of their formation and early evolution we investigate, in this paper, the CB physical conditions at the origin 
given their features at the distance when they become transparent to their 
enclosed radiation as required to explain GRBs.  
Our proposal is that the conditions within the CB as we scale the distance
down along its path to the origin should be an indication of their formation 
site. This, despite the simplicity of our approach, we hope might help elucidate some questions related to the origin/existence of these CBs. We start in Sect. 2 by a brief introduction to the CB model as described in  Dar \& De R{\'u}jula (2004). In particular CBs conditions
 at infinity which best fits GRB lightcurves are isolated. 
In Sect. 3 we present the methods we adopted to extrapolate back to the
  CB source.  In Sect. 4, given the conditions at the origin,
 we study possible formation sites and explore  formation mechanisms. 
Sect. 5 is devoted to the study of mechanisms 
 capable of acceleration CBs to Lorentz factors as high as $\sim 1000$.
 We summarize our results and conclude in Sect. 6.

%-----------------------------------------------------------------------------------
\section{The CB model for GRBs}

In the CB model for GRBs,  the prompt gamma ray emission is
 assumed to be produced when ambient light from the supernova\footnote{The wind from the SN progenitor star is ionized and
is semi-transparent to photons in the visible and UV frequencies.}
 is Compton up scattered by the electrons in the CB. 
These CBs  move with $\Gamma_{\rm CB}\sim1000$
with respect to the supernova remnant and as such
the emitted radiation is highly beamed in the
observer frame (Dar \& De R{\'u}jula 2004).
For the first $10^{3}$\,s in the CB rest frame, the CB in a fast cooling phase 
 emits via thermal bremsstrahlung (Dado et al. 2002), 
but eventually, it is argued, its emissivity is dominated by
 synchrotron emission from ISM electrons that penetrate it. 
A CB will become transparent to the bulk of its enclosed radiation in a time
of ${\cal{O}}(1)$\,s in observer frame after it exits the transparent outskirts
of the shell of the associated SN. 
The internal radiation pressure drops abruptly and its transverse expansion rate
is quenched by collisionless, magnetic-field mediated interactions with the ISM 
(Dado et al. 2002).

Typical values for CB parameters as derived by Dado et al. (2002) and Dar \& De
R{\'u}jula (2004) are given in Table~\ref{typical_CB} where we denote 
the radius of the CB by $R_{\rm CB}$, 
the distance travelled by the CB from its origin by  $D$,  
the time passed in the CB rest frame by $t$.
The expansion velocity of the CB is denoted by $\beta_i=v_i/c$, 
the number of baryons and the mass of the CB by $N_{\rm CB}$ and $M_{\rm CB}$, respectively.
The subscript "trans" refers to the point where the CB becomes transparent ("transparency radius"). 

By fitting the observed GRB {\em afterglow} the CB Lorentz factor is estimated 
to vary between $\Gamma_{\rm CB}=250$ and $\Gamma_{\rm CB}=1600$, while the
number of baryons is of the order of $10^{50}$. With this information at hand,
 our goal is to derive the conditions in the CB as we integrate
 back to a plausible source. As an indication 
 of the close proximity to a compact source (e.g. black-holes and neutron stars)    
when applicable we will make use of the notion of light cylinder
 which we take to be about $R_{\rm L}\sim 10^7$\,cm. 

\begin{table}
\caption{Cannon ball parameters as given by Dado et al. (2002) and Dar \& De R{\'u}jula (2004)}
\centering
\begin{tabular}{ll}
\hline
\noalign{\smallskip}
parameter & value \\
\noalign{\smallskip}
\hline \hline
\noalign{\smallskip}
$R_{\rm CB, max}$ & $2.2\times 10^{14}$ cm \\
$R_{\rm CB, trans}$ & $10^{13}$ cm \\
$D_{\rm trans}$ & $1.7\times 10^{16}$ cm \\
$\Gamma_{\rm CB}$ & $1.0\times 10^3$ \\
$\delta$ & $1.0\times 10^3$ \\
$z$ & $1$ \\
$\beta_{\rm i}$ & $1/\sqrt{3}$ \\
$N_{\rm CB}$ & $10^{50}$ \\
$M_{\rm CB}$ & $10^{26}$ g\\
\noalign{\smallskip}
\hline
\end{tabular}
\label{typical_CB}
\end{table}

%-----------------------------------------------------------------------------------
\section{Cannonball propagation and evolution}
In this section, we will explore the evolution of the CB by extrapolating backward
from the location where the GRB occurs to where the CB reaches nuclear saturation
density (applying CB parameters as given in Table~1). 
For simplicity, we assume that the CB is expanding with a
constant velocity and is moving with a constant Lorentz factor.
The natural assumption for the expansion velocity is the sound speed of the 
hot blob of matter, $v_{\rm exp} \simeq c_s \simeq c/\sqrt{3}$.
Six different cases of CB Lorentz factor and mass are investigated 
(see Table~\ref{cases_CB}).

We will first apply a simple model of the CB's internal energy obeying a simple 
equation of state. 
Using this we calculate the evolution of the density, magnetic field and
temperature as the CB moves away from the origin.
In a second step we extend our model approach applying an energy equation 
where pressure degeneracy and neutrino effects are included. 
As we will see, spatial back integration from the transparency radius will give strong
indication that the 
{\bf CB may be launched close to a black hole.}
The spatial integration back to the source is carried out until the CB temperature reaches
extreme values, $T_{\rm l}\sim 100$ MeV, unless the CB density reaches nuclear saturation
density before $T_{\rm l}$.

With the CB expanding at a constant expansion velocity equal to the sound speed of
the matter $v_{\rm exp}=c/\sqrt{3}$, 
we can use certain estimates about the CB at the distance of transparency to 
derive an interrelation between radius and distance from origin the CB has traveled. 
As radius of the CB at the distance of transparency we apply the estimate by 
Dado et al. (2002),
\begin{equation}
R_{{\rm trans}}\simeq10^{13} \left(\frac{\rm N_{CB}}{6\times 10^{50}}\right)^{1/2} {\rm cm}.
\end{equation}
To reach this radius, the CB has traveled a period of time
\begin{equation}
 t_{{\rm trans, CB}}=\frac{R_{{\rm trans}}}{v_{{\rm exp}}}
\simeq 577 \left(\frac{\rm N_{CB}}{10^{50}}\right)^{1/2}{\rm sec}.
\end{equation}
in the CB rest frame.
As it travels essentially with the speed of light,
 at the time when it becomes transparent, 
the CB has traveled a distance
\begin{equation}
 D_{\rm trans}=\Gamma_{\rm CB}~c~t_{\rm trans}
\end{equation}
from its origin where it was ejected. This gives a linear scaling factor
\begin{equation}
 l=R_{{\rm trans}}/D_{\rm trans}=\frac{1}{\Gamma_{\rm CB} \sqrt{3}}.
\end{equation}

\begin{table}
\caption{The different cases of CBs explored in this work. 
Note that cases 6, 8 and 9 ($\Gamma_{\rm CB}=1000$ and $N_{\rm CB}=10^{51}$, $\Gamma_{\rm CB}=2000$ and $N_{\rm CB}=10^{50}$, $\Gamma_{\rm CB}=2000$ and $N_{\rm CB}=10^{51}$ respectively) are not consistent 
with our assumptions as they reaches nuclear saturation density at a distance beyond
the light cylinder, and have therefore been left out in this paper. }
\centering
\begin{tabular}{lll}
\hline
\noalign{\smallskip}
 & $\Gamma_{\rm CB}$ & $N_{\rm CB}$ \\
\noalign{\smallskip}
\hline \hline
\noalign{\smallskip}
Case 1 & $1.0\times10^2$ & $10^{49}$ \\
Case 2 & $1.0\times10^2$ & $10^{50}$ \\ 
Case 3 & $1.0\times10^2$ & $10^{51}$ \\
\noalign{\smallskip}
\hline			   
\noalign{\smallskip}
Case 4 & $1.0\times10^3$ & $10^{49}$ \\
Case 5 & $1.0\times10^3$ & $10^{50}$ \\ 
\noalign{\smallskip}
\hline			   
\noalign{\smallskip}
Case 7 & $2.0\times10^3$ & $10^{49}$ \\
\noalign{\smallskip}
\hline
\end{tabular}
\label{cases_CB}
\end{table}

\begin{table}
\caption{The CB radius at the point where it approaches density reaches nuclear saturation density
({\em left}), at the light cylinder $R_{\rm LC}=1.5\times10^7$ cm ({\em middle}) and 
at the distance where the CB becomes transparent to radiation ({\em right}).}
\centering
\begin{tabular}{llll}
\hline
\noalign{\smallskip}
 & $ R_{\rm CB, nuc} [cm]$ & $R_{\rm CB, LC} {\rm [cm]}$  & $R_{\rm CB, trans} [cm]$ \\
\noalign{\smallskip}
\hline \hline
\noalign{\smallskip}
Case 1  & $2.5\times10^{3}$  & $1.2\times10^{5}$   & $1.0\times10^{12}$ \\
Case 2  & $5.6\times10^{3}$  & $1.2\times10^{5}$   & $1.0\times10^{13}$ \\ 
Case 3  & $12.0\times10^{3}$ & $1.2\times10^{5}$   & $1.0\times10^{14}$ \\
\noalign{\smallskip}
\hline 
\noalign{\smallskip}
Case 4  & $2.5\times10^{3}$  &  $1.2\times10^{4}$  & $1.0\times10^{12}$ \\
Case 5  & $5.6\times10^{3}$ &  $1.2\times10^{4}$  & $1.0\times10^{13}$ \\ 
\noalign{\smallskip}
\hline
\noalign{\smallskip}
Case 7  & $2.5\times10^{3}$ & $5.8\times10^{3}$  & $1.0\times10^{12}$ \\
\noalign{\smallskip}
\hline
\end{tabular}
\label{linear_CB}
\end{table}

The radius of a CB is then simply expressed as $R_{\rm CB}=D/\Gamma_{\rm CB}\sqrt{3}$. 
The radius at nuclear saturation density and at the light cylinder ($D_{\rm lc}\sim 1.5\times10^7$ cm) 
are listed in Table~\ref{linear_CB} as well as the radius at which the CB becomes transparent to the
enclosed radiation for the different cases of CB masses and Lorentz factors. 

\begin{figure}[h!]
\includegraphics[scale=0.7]{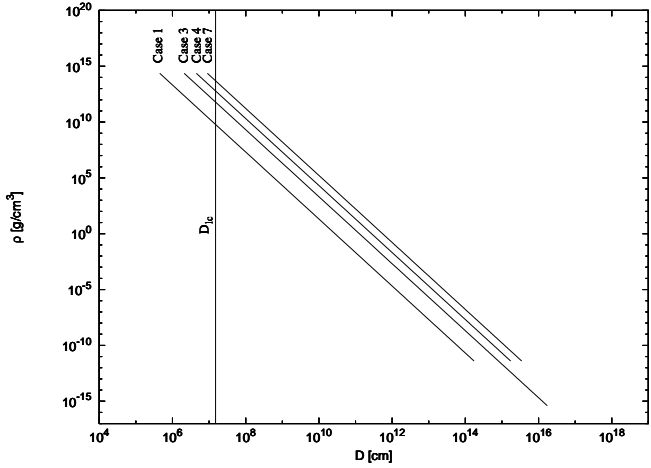}
\caption{Density vs distance from the origin for the CB.
 The backward integration is stopped when the CB density reaches nuclear saturation density.}
\label{density}
\end{figure}

Figure \ref{density} shows density as a function of
distance travelled by the CB. 
Depending on the CB parameters, the densities at the light cylinder range from $10^{10}-10^{14} {\rm g/cm^3}$.

%--------------------------------------------------------------  Fig 1 

%--------------------------------------------------------------  Fig 2 
\begin{figure}[h!]
\includegraphics[scale=0.7]{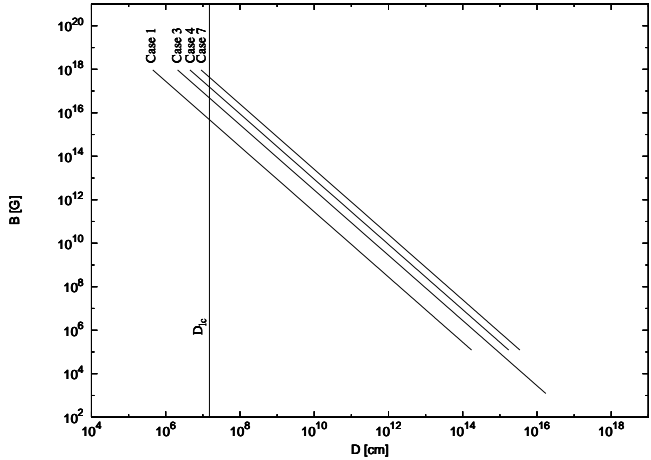}
\caption{Magnetic field vs distance from the origin for the CB. 
The backward integration is stopped when the CB density reaches nuclear saturation density.}
\label{magfield}
\end{figure}

If we estimate the CB magnetic field strength from the equipartition assumption,
$v_{\rm A}\simeq v_{\rm s}=c/\sqrt{3}$, with the Alfv\'en speed $v_{\rm A}$ and
\begin{equation}
 B=v_{\rm A}\sqrt{4\pi\rho}=c\sqrt{\frac{M_{\rm CB}}{R_{\rm CB}^3}}.
\label{magfieldeq}
\end{equation}
we may compute the CB magnetic field versus the distance from origin (Fig.~\ref{magfield}).
Close to the hypothetical CB origin where we reach nuclear saturation density, 
the magnetic field strength approaches values up to $B\sim 10^{18} G$ (see also \S \ref{accdisks}).
At a distance from the origin of the order of the light cylinder radius,
the field strength is between $4.8\times10^{15}$ and
$1.4\times10^{18}$ Gauss depending on the choice of Lorentz factor and mass of the CB. The densities ranges from $10^9$ to $10^{14} {\rm g/cm^3}$. 

%------------------------------------------------------------------------------------------
\subsection{Simple energy equation}
We continue our simple estimates by assuming energy conservation in the CB
\begin{equation}
 E_{\rm rad}+E_{\rm th}+E_{\rm mag}=E_{\rm tot}.
\label{energyeqnn}
\end{equation}
The radiation energy is written as,
\begin{equation}
 E_{\rm rad}=aT^4 \frac{4}{3}\pi R_{\rm CB}^3\ ,
\label{radene}
\end{equation}
the magnetic energy is,
\begin{equation}
E_{\rm mag}= \epsilon_{\rm m} M_{\rm CB}c^2
\label{magneticene}
\end{equation}
and the gas thermal energy is,
\begin{equation}
 E_{\rm th}=3N_{\rm CB}kT,
\label{gasene}
\end{equation}
where $a=7.5657\times10^{-15}{\rm \,erg\, cm^{-3}\, K^{-4}}$ is the radiation constant, $k=1.3807\times10^{-16} {\rm \,erg\, K^{-1}}$ is Boltzmann's constant, $T$ is the CB temperature, $\epsilon_{\rm m}$ is a parameter that allow us to write the magnetic energy in terms of the CB rest mass energy. This parameter is fixed by imposing energy equipartition at the specified CB origin.
Note that the magnetic energy is constant, as we assume that the magnetic field is not dissipated in reconnection events and it is not expelled from the CB.
The gravitational energy is always negligible compared to the other energy channels. 
The total energy is then given as:
\begin{equation}
E_{\rm tot}=3\times E_{\rm mag}=3\epsilon_{\rm m}M_{\rm CB}c^2.
\end{equation}
We can now write the energy equation as:
\begin{equation}
4/3\pi R^3 aT^4+3NkT=2\epsilon_{\rm m} Mc^2
\label{totene}
\end{equation}
This equation is solved to find $T$ as a function of $D$.

In what follows, we explore two scenarios: (i) the first
 one is indicative of the close proximity of a compact source and
 as such it corresponds to the case where the energy equation is
integrated assuming equipartition at nuclear saturation densities;
 (ii) the second reflect scenarios where the CB originates
 from the coronal region of compact stars of of their associated accretion
disks. Specifics below.

\subsubsection{Equipartition at nuclear densities: {\it source origin}}
\begin{figure}[h!]
\includegraphics[scale=0.7]{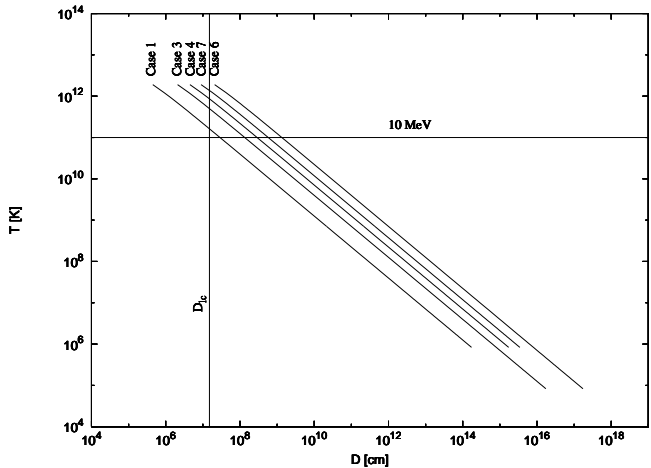}
\caption{Temperature vs distance traveled by the CB assuming equipartition at nuclear saturation density. }
\label{t4s}
\end{figure}

\begin{figure}[h!]
\includegraphics[scale=0.7]{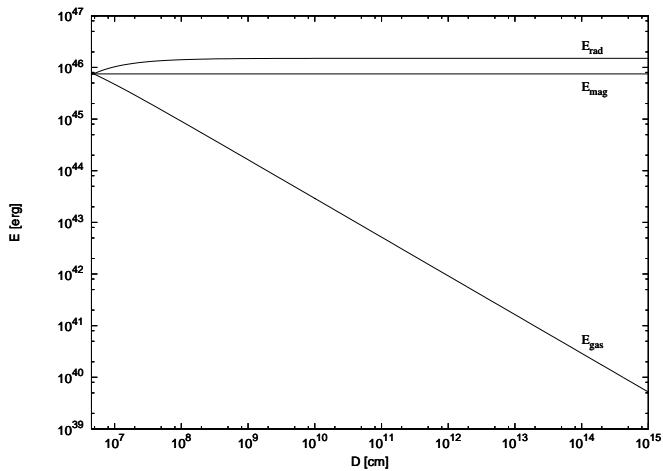}
\caption{The different energy components for case 4 using the simple energy equation and assuming equipartition at nuclear densities.}
\label{e4s}
\end{figure}

To set equipartition at nuclear saturation density, $\epsilon_m$ has to be $0.5$ for all cases.
By rearranging Eq. (\ref{totene}) it can be seen that it becomes a function 
of T and $\rho$:
\begin{equation}
aT^4+3\rho kT/m_H=2 \rho c^2 \epsilon_{\rm m}
\end{equation}
The temperature therefore depends on the density only and implies
a temperature of about $10^{12}$ K at nuclear saturation density for all 
cases. The temperature as a function of distance travelled is shown in Fig.~\ref{t4s} while
Fig.~\ref{e4s} shows the energy components for case 4. The radiation energy 
is dominant everywhere except
 at nuclear saturation density where there is equipartition.
We note that nuclear saturation density would be reach at distances larger than $10^6$ cm for most cases. It is unrealistic to find object with such high densities much larger than $10^6$ cm. Also, CBs with such densities need a magnetic field $B=10^{18}$ G, which is unrealisticly high. We can therefore rule out CBs formed with nuclear saturation density.

It should be noted that the solutions are not very sensitive to the choice of 
total internal energy\footnote{The expansion
energy of the CB is of the same order as the magnetic energy. For simplicity 
we have included it
in the expression for the total energy.}. As an example, increasing the total 
internal energy by an order of magnitude we find a temperature at the light 
cylinder, $D_{\rm lc}$,
 for case 3 to be $9\times 10^{11}$ K, compared to $5\times10^{11}$ K in our 
initial calculation.

\subsubsection{Equipartition at the light cylinder: {\it coronal origin}}
\begin{figure}[h!]
\includegraphics[scale=0.7]{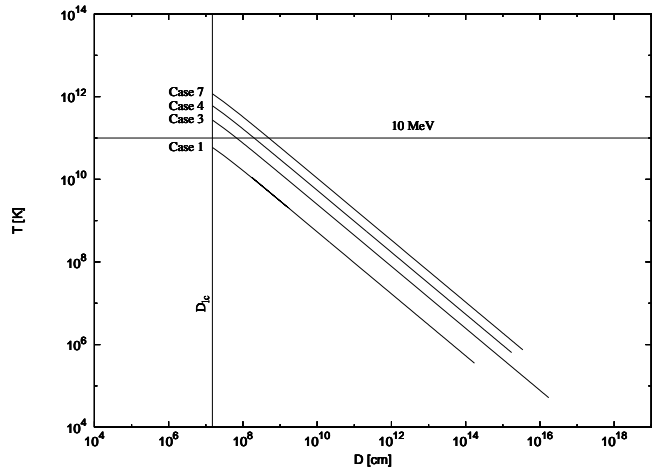}
\caption{Temperature vs distance traveled by the CB assuming equipartition at the light cylinder. }
\label{t4s_lc}
\end{figure}

\begin{figure}[h!]
\includegraphics[scale=0.7]{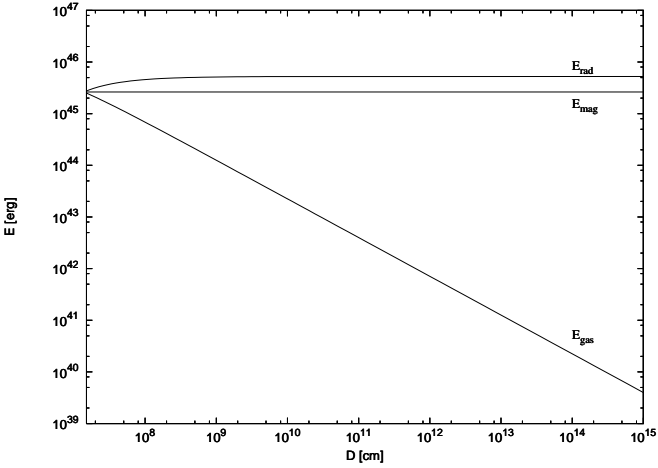}
\caption{The different energy components for case 4 using the simple energy equation and assuming
 equipartition at the light cylinder.}
\label{e4s_lc}
\end{figure}

Figure \ref{t4s_lc} show the evolution
 of the CB conditions when 
 equipartition at a distance of about a light cylinder radius is assumed, 
and in Fig.~\ref{e4s_lc} the energy components for case 4.
Table \ref{a1res_lc} shows the temperature found at the light cylinder and the value for the energy equipartition parameter $\epsilon_{\rm m}$ 
for different kinematic parameters of the CB (see Table\ref{typical_CB}).
Note that $\epsilon_{\rm m}$ is now {\em determined} by the condition that we have 
equipartition at the CB origin (i.e. at a light cylinder distance). 

\begin{table}[h!]
\caption{Coronal CB origin. The values for the temperature and $\epsilon_{\rm m}$ found for the
simple energy equation assuming equipartition between the magnetic, gas thermal and radiation
energy at the light cylinder. }
\centering
\begin{tabular}{lll}
\hline
\noalign{\smallskip}
Case & $T$ [K] & $\epsilon_{\rm m}$ \\
\noalign{\smallskip}
\hline \hline
\noalign{\smallskip}
1 & $5.9\times10^{10}$ & $0.0165$ \\
2 & $1.3\times10^{11}$ & $0.035$ \\ 
3 & $2.7\times10^{11}$ & $0.075$ \\ 
4 & $6.1\times10^{11}$ & $0.175$ \\ 
5 & $1.2\times10^{12}$ & $0.325$ \\ 
7 & $1.2\times10^{12}$ & $0.325$ \\ 
\noalign{\smallskip}
\hline
\end{tabular}
\label{a1res_lc}
\end{table}

By assuming an equal number of electrons $n_e$ and baryons in the CB the Fermi 
temperature can be computed.
If we take the light cylinder distance as a typical length unit at the CB origin and compute 
the Fermi temperature at this location, we see that the CB temperature is much smaller than the 
electron Fermi temperature (Table~\ref{a1res}).
We therefore have to improve our approach considering also electron degeneracy and neutrino effects.
In the next section we will explore a more appropriate energy equation where neutrino effects
are added. 
However, we will still apply a kinematic approach assuming a linear expansion of the CB.

\begin{table}[h!]
\caption{The values for the magnetic field, temperature and density found at the light cylinder for
the simple energy equation. The temperature and magnetic field at nuclear saturation density
is $1.2\times10^{12}$ K and $9.3\times10^{17}$ G respectively for all cases, as both the magnetic
field strength and temperature is dependent on the density only.}
\begin{tabular}{lllll}
\hline
\noalign{\smallskip}
Case & $B$ [G] & $T$ [K] & $T_{\rm F}$ [K] & $\rho$ [${\rm g cm}^{-3}$]  \\
\noalign{\smallskip}
\hline \hline
\noalign{\smallskip}
1 & $4.8\times10^{15}$ &
$5.9\times10^{10}$ & $5.6\times10^{14}$ & $6.1\times10^{9}$ \\
2 & $1.5\times10^{16}$ & 				    
$1.3\times10^{11}$ & $2.6\times10^{15}$ & $6.1\times10^{10}$ \\
3 & $4.8\times10^{16}$ &				    
$2.7\times10^{11}$ & $1.2\times10^{16}$ & $6.1\times10^{11}$ \\
4 & $1.5\times10^{17}$ &				    
$6.0\times10^{11}$ & $5.6\times10^{16}$ & $6.1\times10^{12}$ \\
5 & $4.8\times10^{17}$ &				    
$1.2\times10^{12}$ & $2.6\times10^{17}$ & $6.1\times10^{13}$ \\
7 & $4.3\times10^{17}$ &				    
$1.2\times10^{11}$ & $2.3\times10^{17}$ & $4.9\times10^{13}$ \\
\hline
\end{tabular}
\label{a1res}
\end{table}

\subsection{Coronal CB origin: {\it neutrino effects}}
In this section we improve our approach 
 by taking into account degeneracy pressure
and neutrino cooling (e.g. Popham et al. 1999). The energy
equation becomes,
\begin{equation}
 E_{\rm tot}+E_{\rm \nu}(t)=E_{\rm th}+E_{\rm rad}+E_{\rm deg}+E_{\rm mag}
\label{degtot}
\end{equation} 
where 
\begin{equation}
E_{\rm deg} = 3KM_{\rm CB}\bigg(\frac{\rho}{\mu_{\rm e}}\bigg)^{1/3}\ ,
\label{degene}
\end{equation}
is the degeneracy energy,
\begin{equation}
 E_{\rm rad}=\frac{11}{4}\frac{aT^4}{\rho} M_{\rm CB},
\label{raddegene}
\end{equation}
is the radiation energy and
\begin{equation}
 E_{\rm th}=\frac{3}{2}RTM_{\rm CB}\frac{1+3X_{\rm nuc}}{4},
\label{gasdegene}
\end{equation}
the gas thermal energy where
\begin{equation}
\begin{split}
X_{\rm nuc}=30.97\bigg(\frac{\rho}{10^{10} {\rm g/cm^3}}\bigg)^{-3/4}\bigg(\frac{T}{10^{10} {\rm K}}\bigg)^{9/8} \\
\times \exp\bigg(-6.096\times\frac{10^{10}{\rm K}}{T}\bigg)
\end{split}
\end{equation}
gives $X_{\rm nuc}<1$, and $X_{\rm nuc}=1$ elsewhere.
In the equation above, $K=(2\pi h c/3)(3/8\pi m_n)^{4/3} = 1.24 \times 10^{15}$, $m_n$ is the nucleon 
mass, R is the gas constant, a is the radiation constant and $\mu_e=2$ is the mass per electron.
Inserting Eqs. (\ref{magneticene}), (\ref{degene}), 
(\ref{raddegene}) and (\ref{gasdegene}) into (\ref{degtot}) gives in terms of $D$
\begin{equation}
\begin{split}
E_{\rm tot} & +E_\nu(t) = 3KM_{\rm CB}\bigg(\frac{M_{\rm CB} 3^{5/2}}{4\pi\mu_{\rm e}}\bigg)^{1/3}\frac{\Gamma_{\rm CB}}{D}\\ 
&+\frac{3}{2}RTM_{\rm CB}\frac{1+3X_{\rm nuc}}{4}
 +\frac{11}{4}aT^4 \frac{4\pi D^3}{\Gamma_{\rm CB}^3 3^{5/2}}
\end{split}
\label{pophamenergyD}
\end{equation}

Note that we apply the same total internal energy ($E_{\rm tot}$) of the CB as in Sec. 3.3,
thus the same energy parameter $\epsilon_{\rm m}$.
Close to the origin, however, we add an energy component due to neutrino effects
 (emissivity and cooling) and is denoted   by $E_\nu(t)$ in Eq.~\ref{degtot}.

Two types of neutrino losses may occur, 
i.e. neutrino emission due to pair annihilation and 
neutrino losses due to the capture of pairs on nuclei.
Their contribution to the energy budget is computed from Eq. 3.8 and Eq. 3.9 in Popham et al. (1999):
\begin{equation}
\dot{q}_{\nu \overline{\nu}}=5.0\times 10^{33} \bigg(\frac{T}{10^{11} K}\bigg)^9 {\rm ergs \, cm^{-3}\, s^{-1}}
\end{equation}
\begin{equation}
\dot{q}_{eN}=9.0\times10^{33}\bigg(\frac{\rho}{10^{10} {\rm g/cm^3}}\bigg)\bigg(\frac{T}{10^{11} K}\bigg)^6 {\rm ergs \, cm^{-3}\, s^{-1}}
\label{neutrinocooling}
\end{equation}
These expressions are integrated over the time it takes the CB to reach conditions for which 
neutrino cooling is not significant. We find the latter cooling method (Eq.~\ref{neutrinocooling}) to be dominant, so we limit ourself to using that.

To compute the effects due to neutrinos, we must know the temperature.
However, in turn, we want to use the neutrino effects to find the temperature.
We therefore first solve the energy equation \ref{degtot} without adding neutrino effects. Then we use the temperature found to calculate the neutrino emissivity which is then added to the total energy in equation \ref{degtot}, and this equation is solved to find the temperature as a function of distance travelled. The neutrinos are released in small successive bursts, mimicking a continuous emission.

We emphasize again that because of the neutrinos effects and the different energy equation used
in sec 3.3 there is no assumption on energy equipartition applied in this section.

\begin{figure}[h!]
\includegraphics[scale=0.7]{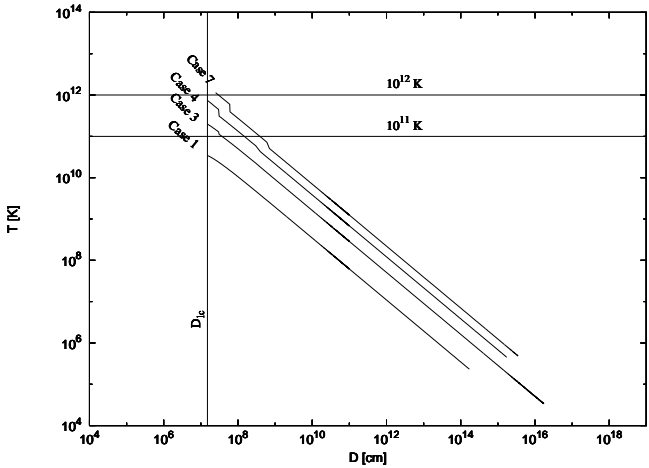}
\caption{The temperature assuming the same total energy as in sec. 3.3 with neutrino and degeneracy effects. Case 7 goes back to $10^{12}$ K (at $D=2.5\times 10^{12}$ cm), whereas case 1, 3 and 4 goes to the light cylinder.}
\label{t4lcnu}
\end{figure}

\begin{figure}[h!]
\includegraphics[scale=0.7]{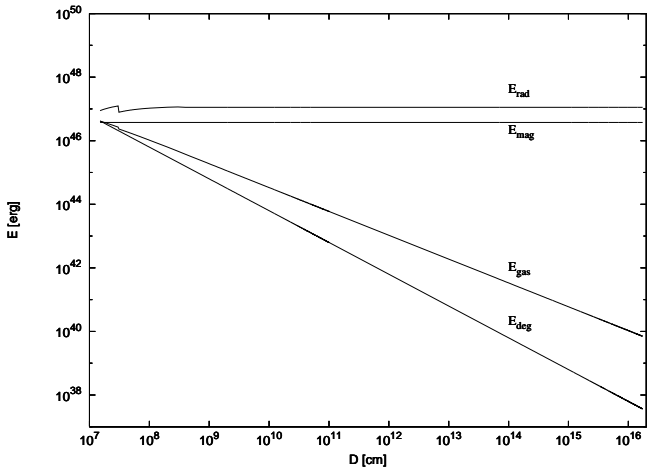}
\caption{The energy components for case 3, including degeneracy and neutrino effects and assuming that case 3 originate at the light cylinder. The neutrino release is visible in the radiation energy and gas thermal energy at $D=3\times10^7$ cm. Another not visible release occurs at $D=3\times10^8$ cm.}
\label{e3enelcnu}
\end{figure}

The temperature is shown in Fig.~\ref{t4lcnu} as a function of distance travelled. 
In general, the release of neutrinos are seen as a small jump in the temperature curve.
We note that cases 1 to 4 reach the light cylinder with reasonable temperatures ($T<10^{12}$ K) and densities ($\rho<10^{14} {\rm g/cm^3}$).
Cases 5 and 7 are ruled out as they reach even more extreme conditions before reaching the light cylinder.
For illustrative purposes in Fig.~\ref{e3enelcnu} we show the energy components for case 3. Because of the neutrino effects, the radiation energy is now the dominant energy, even close to the light cylinder.

%-----------------------------------------------------------------------------------
\section{Sites and formation mechanisms}

In this section we discuss sites
 that are best suited to account for the CBs
 conditions at the source derived in our previous section. 
 We also explore possible formation scenarios.

%---------------------------------------------------------------------------
\subsection{``Standard" and hyperaccreting disks}
\label{accdisks}

In appendix \ref{appendixdisks} we summarize properties of "standard" ($\alpha$-disks
 and advection dominated accretion flow disks) and hyperaccretion disks.
 It is clear that ``standard" accretion disks are ruled out.
Hyperaccreting disks on the other hand are candidates as CB sources. 
They have densities, temperatures and magnetic field that are comparable to what we found in the previous section. In fact when we  consider CBs with hyperaccretion disk conditions at the source and perform a {\it forward} integration, the conditions at $D_{\rm trans}$ (see Figures \ref{forwardT}-\ref{rhoforward} in appendix \ref{appendixforward}) turn out to be interestingly similar to those given by Dado et al. (2002). The next step then is to look for formation mechanisms within the hyperaccreting context.

The stability of hyperaccreting disks around black holes have been recently
investigated by Di Matteo et al. (2002) who find that the ``flows are gravitationally 
stable under almost all conditions of interest''.
Exceptions exist for strong accretion rates and in the outer part of the disk
(see also Narayan et al. 2001). However as can be seen
for Eq. (\ref{hyperrho})-(\ref{hyperb}) in the appendix these extreme cases favor
lower densities and temperatures than those expected for CBs.

The magneto-rotational instability (MRI, Balbus \& Hawley 1991 and Hawley \& Balbus 1991) 
works only for low magnetic field strengths and cannot account for
the strong magnetic fields required at the origin for CBs.
Let us also mention the {\em accretion-ejection} instability (Tagger et al. 1992) 
as a possible formation mechanism. This instability works for intermediate magnetic field strengths and will transfer angular momentum to Alfv{\'e}n waves toward the corona of the disk.  At extreme magnetization 
 the {\em accretion-ejection} instability is reminiscent of
 the {\em interchange instability} (Spruit et al. 1995) but it seems unlikely that these can lead to CBs formation since most of the perturbations
 are carried by Alfv\'en waves.

 It is thus not clear how a CB can form within a hyperaccretion disk. There is also the issue of accelerating the CB to $\Gamma>100$ which is also a major challenge. We will return to this in \S \ref{sectionacceleration} after we discuss other possible formation sites.

\subsection{Neutron tori}

The thick, self-gravitating, neutron tori around 2-3 $\msun$
black holes are known to be affected by a runaway instability on time scales below 
the evolutionary time scale of GRBs (Nishida \& Eriguchi 1996) and we
 therefore exclude them as source for CBs.
Simulations of neutron star mergers have also shown that about $0.01\,\msun$ of the
thick disk of $0.2\,\msun$ around a 1.5 to 3.1 $\msun$ final central mass distribution
becomes gravitationally unbound (Rosswog et al. 1999).
However, in difference to the hyperaccreting disk model, this unbound mass stays
rather cold ($10^8 K$) and do not constitute a formation site for CBs.

%---------------------------------------------------------------------------
\subsection{Accretion disk corona}

Another possibility is CB formation in the disk corona, for example
  as a huge magnetic flare which 
ejects a large part of the accretion disk corona into a bullet of high velocity.
A CB of such a size would have a density of about 
$2.4 \times 10^5\,{\rm g\,cm^{-3}}$ which is, for comparison, in the range
of white dwarf densities.
The maximum initial size of the CB we expect not to exceed $\rl \simeq 10^7$\,cm.
Comparing the CB asymptotic kinetic energy to the magnetic energy contained in a 
volume of that size, provides an estimate for the mean magnetic field strength
of about $10^{15}$G.
This corresponds to a magnetic flux of $\sim10^{29}-10^{30}{\rm G\,cm^2}$ and is unrealistically high for such coronae.

\subsection{Disk-jets and funnel-jets}

\begin{figure}[h!]
\includegraphics[scale=0.5]{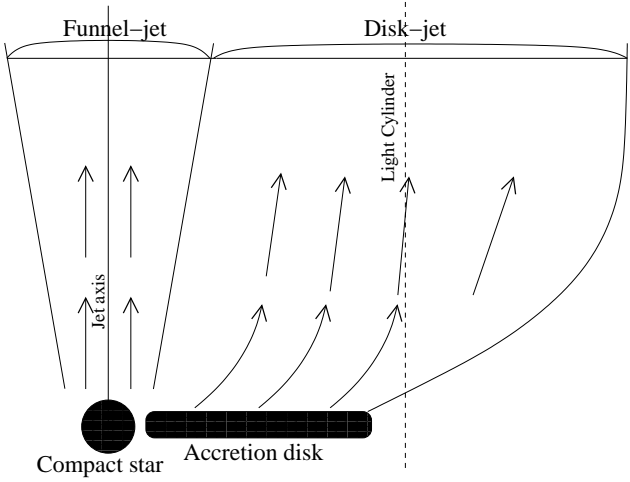}
\caption{Illustration of Funnel-jet and Disk-jet. The funnel-jet is launched from a region close to the compact star. The disk-jet is launched from the accretion disk.}
\label{jets}
\end{figure}

Figure \ref{jets} is an illustration of the type of jets
that could emanate from the vicinity of a compact star.
 The disk-jet material is ejected from the accretion disk
 while the funnel-jet is ejected from the innermost parts
 of the disk at the interface with the compact star.

Recent general relativistic magneto-hydrodynamic simulations by De Villiers et al. (2005) of a black hole and an initial torus seeded with a weak poloidal magnetic field show that a funnel jet with $\Gamma_{\rm CB} \gtrsim 50$ is formed. Instabilities do occur in funnel-jets, however, the induced instabilities have densities much lower than the CB values found in the previous section. Funnel-jets can therefore be ruled out as a possible formation site for CBs. 

A disk-jet becomes cylindrically collimated on a length scale of the
order of 1-2 light cylinder distances (Fendt \& Memola
2001). Knot generating instabilities reminiscent of CBs are known to occur as jets collimate (Ouyed et al. 1997). 
This is a possible formation mechanism for CBs. What remains is to show how they can be accelerated to high $\Gamma$. Specifics below.

%---------------------------------------------------------------------------
\section{CB acceleration to ultra-relativistic velocity}
\label{sectionacceleration}

Having isolated, or more precisely favored, jets from hyperaccretion disks as plausible
 formation sites for CBs we now discuss acceleration mechanisms with which
 CBs can reach Lorentz factors in the thousands.

Assuming that the CB is accelerated by converting the internal magnetic energy to kinetic energy, we can find an estimate for the magnetic field needed to explain such Lorentz factors. Using typical CB radii close to the source and parameters from Table~\ref{typical_CB}
we find that the magnetic field must be of the order of $10^{18}-10^{19}$ G. This is unrealistically high. Either our approach is too simple, or a different acceleration mechanism must be at work. 

\subsection{MHD Acceleration: CB speed}

The ability of the magnetic field to accelerate particles to high Lorentz factors is given by the magnetization parameter (Michel 1969)
\begin{equation}
\sigma=\frac{\Phi^2\Omega_F^2}{4\dot{M}_{\rm jet}c^3},
\end{equation}
where $\Phi=B_pr^2$ is the magnetic flux, $\Omega_F=c/D_{\rm lc}$ is the angular frequency of the magnetic field and $\dot{M}_{\rm jet}=\pi\rho v_pr^2$ is the mass flow rate within the flux surface. For spherical outflow Michel (1969) found that the Lorentz factor at infinity scales as
\begin{equation}
\Gamma_\infty=\sigma^{1/3}.
\end{equation}

Fendt \& Ouyed (2004) finds a modified Michel scaling in the case of a non-spherical magnetic field distribution. In this case they find a linear relation between $\sigma$ and $\Gamma_\infty$. If the field distribution is $\Phi(r;\Psi)\sim r^{-0.1}$, they find that $\Gamma_\infty=10^{-1/3}\sigma$, and if $\Phi(r;\Psi)\sim r^{-0.2}$, they find 
\begin{equation}
\Gamma_\infty=10^{-1/5}\sigma\ , 
\end{equation}
in which case hyperaccreting disks with ejection rates 
 of the order $10^{-5}M_\odot/s$ and magnetic field of the order $10^{14}G$ can  lead to jets with a Lorentz factor $\Gamma_\infty\simeq 1875$. 

\subsection{MHD instability: CB mass}

To a first order, instabilities related to Alfv\'en crossing time
 can develop on timescales
\begin{equation}
t_{\rm ins}=t_{\rm A}=\frac{2R_{\rm jet}}{v_{\rm A}},
\end{equation}
where $R_{\rm jet}$ is the radius of the disk-jet. For $1 R_{\rm lc}<R_{\rm jet}<10R_{\rm lc}$, we arrive at $t_{\rm ins}\sim 1-10$ ms which would imply
 the plausible formation of blob of matter as massive as
 $M_{\rm ins}=t_{\rm ins}\dot{M}_{\rm jet}\sim10^{-8}-10^{-7} M_\odot$. This can be compared to the typical CB mass of the order $M_{\rm CB}=10^{-7}M_\odot$.

As we have shown above, first forming the CB in the disk and then accelerating it will require unrealistic magnetic fields of the order $10^{19}$ G. However, first accelerating the wind to the light cylinder {\it and then} forming the CB through an instability beyond the light cylinder requires much smaller magnetic field strength ($<10^{14}$ G). This is a possible mechanism for forming and accelerating CBs.

%---------------------------------------------------------------------------------
\section{Conclusion}

Assuming that CBs move and expand with a constant velocity we have estimated the CB conditions as close as possible to their origin.
CBs require extremely high internal magnetic fields when they are formed
with field strength exceeding $\sim 10^{15}$ G. 
The temperature was found to
be of the order of $10^{11} - 10^{12}$ K. 
The physical parameters of the CBs at the origin are,
 within an order of magnitude estimates, indicative of hyperaccreting disks. 
 However, if formed in the accretion disk we find
 it challenging to accelerate the CBs to the high Lorentz factors.  
The coronal origin is ruled out because of the unrealistically 
high coronal magnetic flux necessary to form the CBs.
Our results instead hint to a jet origin of CBs. 
The radius ($< D_{\rm lc}$) and mass flow ( $10^{-5} M_\odot/s$) in a jet from a hyperaccreting can account for the CB mass and density.
 Furthermore, this outflow can be accelerated to $\Gamma\sim 2000$ by MHD processes
(Fendt \& Ouyed 2004). Any instability in this outflow beyond the light cylinder could lead to CB formation.  We thus suggest that CBs form as instabilities in ultra-relativistic jets emanating from the surface of hyperaccretion disks. 
The tight link between SNe and the CB model for GRB  
  requires that all (or almost all) core collapse SNe will produce CBs. 
Our work, within its limitations, implies that hyperaccretion disks must be a common occurrence 
 in core collapse SNe to accommodate the CB model - a notion which remains to be confirmed.

\begin{acknowledgements}
The research of R. O. is supported by an operating grant from the Natural Science and Engineering Research Council of Canada (NSERC) as well as the Alberta Ingenuity Fund (AIF). J.S. thanks the Canadian Institute for Theoretical Astrophysics for hospitality. C.F. thanks R.O. for the gracious hospitality
at the University of Calgary when much of this work was completed.
\end{acknowledgements}

%---------------------------------------------------------------------------------

%---------------------------------------------------------------------------
\appendix
\section{Accretion disks}
\label{appendixdisks}

\subsection{``Standard" accretion disks}
A standard Shakura-Sunyaev disk (Shakura \& Sunyaev 1973) will have a
density
\begin{equation}
\begin{split}
 \rho[{\rm g\, cm}^{-3}]=7.2\times10^{-4}\bigg(\frac{\alpha_{\rm v}}{0.001}\bigg)^{-1}
\bigg(\frac{\dot{M}}{M_{{\rm Edd}}}\bigg)^{-2} \\
\bigg(\frac{r}{3r_{\rm S}}\bigg)^{3/2} 
\times\bigg(\frac{M}{M_\odot}\bigg)^{-1}\bigg(1-\bigg(\frac{r}{3r_{\rm S}}\bigg)^{-1/2}\bigg)^{-2}.
\end{split}
\end{equation}
where $\alpha_{\rm v}$ is a viscosity parameter, $r_S$ is Schwarzschild radius and
$M_{\rm Edd}$ is the Eddington mass.
With this density, the radius of a CB with mass $M=10^{50}$ baryons becomes
$3.8\times10^9$ cm, assuming the default parameters.
For the equipartition magnetic field one gets
\begin{equation}
B[{\rm G}]=10^8\bigg(\frac{M}{M_\odot}\bigg)^{-1/2}\bigg(\frac{r}{3r_{\rm S}}\bigg)^{-3/4}.
\end{equation}
The temperature is
\begin{equation}
 T[{\rm K}]=1.3\times10^8\bigg(\frac{\alpha_{\rm v}}{0.001}\bigg)^{-1/4}
\bigg(\frac{M}{M_\odot}\bigg)^{-1/4}\bigg(\frac{r}{r_{\rm S}}\bigg)^{-3/4}.
\end{equation}

Advection Dominated Accretion Flow (ADAF) disks have density (Narayan
\& Yi 1995):
\begin{equation}
\begin{split}
 \rho[{\rm
g\, cm}^{-3}]=6.5\times10^{-3}\bigg(\frac{\alpha_{\rm v}}{0.001}\bigg)^{-1}c_1^{-1}c_3^{-1/2} \\
\bigg(\frac{\dot{M}}{M_{\rm Edd}}\bigg)^{+1} 
\times\bigg(\frac{M}{M_\odot}\bigg)^{-1}\bigg(\frac{r}{3r_{\rm S}}\bigg)^{-3/2},
\end{split}
\end{equation}
where $c_1$ and $c_3$ are defined in Eq. (2.1) in Narayan \& Yi
(1995).

The corresponding magnetic field is 
\begin{equation}
\begin{split}
 B[{\rm G}]=5.5\times
10^9\bigg(\frac{\alpha_{\rm
v}}{0.001}\bigg)^{-1/2}c_1^{-1/2}c_3^{1/4}(1-\beta)^{1/2}\\
\times\bigg(\frac{\dot{M}}{M_{\rm Edd}}\bigg)^{1/2}
\bigg(\frac{M}{M_\odot}\bigg)^{-1/2}\bigg(\frac{r}{3r_{\rm S}}\bigg)^{-5/4},
\end{split}
\end{equation}

while the ion temperature of such disks are (Narayan et al. 1998)
\begin{equation}
T_{\rm i}[K] = 2\times10^{12}\beta \bigg(\frac{r}{2r_{\rm s}}\bigg)^{-1},
\end{equation}
where $\beta$ is given by 
\begin{equation}
p_{\rm m}=\frac{B^2}{24\pi}=(1-\beta)\rho c_{\rm s}^2.
\end{equation}

%---------------------------------------------------------------------------
\subsection{Hyperaccreting disks}
Hyperaccreting disk (Popham et al. 1999) density is:
\begin{equation}
\begin{split}
 \rho[{\rm g\,cm}^{-3}]=1.3\times
10^{12}\bigg(\frac{\alpha_{\rm v}}{1.0}\bigg)^{-1.3}\bigg(\frac{\dot{M}}{M_\odot
s^{-1}}\bigg)^{+1}
\\ \times\bigg(\frac{M}{M_\odot}\bigg)^{-1.7}\bigg(\frac{r}{3r_{\rm
S}}\bigg)^{-2.55},
\end{split}
\label{hyperrho}
\end{equation}
their disk scale height is
\begin{equation}
 H[{\rm cm}]
= 1.9\times10^5
\bigg(\frac{\alpha_{\rm v}}{1.0}\bigg)^{0.1}\bigg(\frac{M}{M_\odot}\bigg)^{0.9}
\bigg(\frac{r}{3r_{\rm S}}\bigg)^{1.35},
\end{equation}
while the temperature is  
\begin{equation}
 T[{\rm K}]
= 7.6\times 10^{10}\bigg(\frac{\alpha_{\rm v}}{1.0}\bigg)^{0.2}
\bigg(\frac{M}{M_\odot}\bigg)^{-0.2}\bigg(\frac{r}{3r_{\rm S}}\bigg)^{-0.3}.
\end{equation}
The corresponding equipartition magnetic field is of the order of
\begin{equation}
 B[{\rm G}]\sim 10^{14}-10^{15}.
\label{hyperb}
\end{equation}

\section{Forward integration}
\label{appendixforward}

%--------------------------------------------------------------  Fig B1 
\begin{figure}[b!]
\includegraphics[scale=0.7]{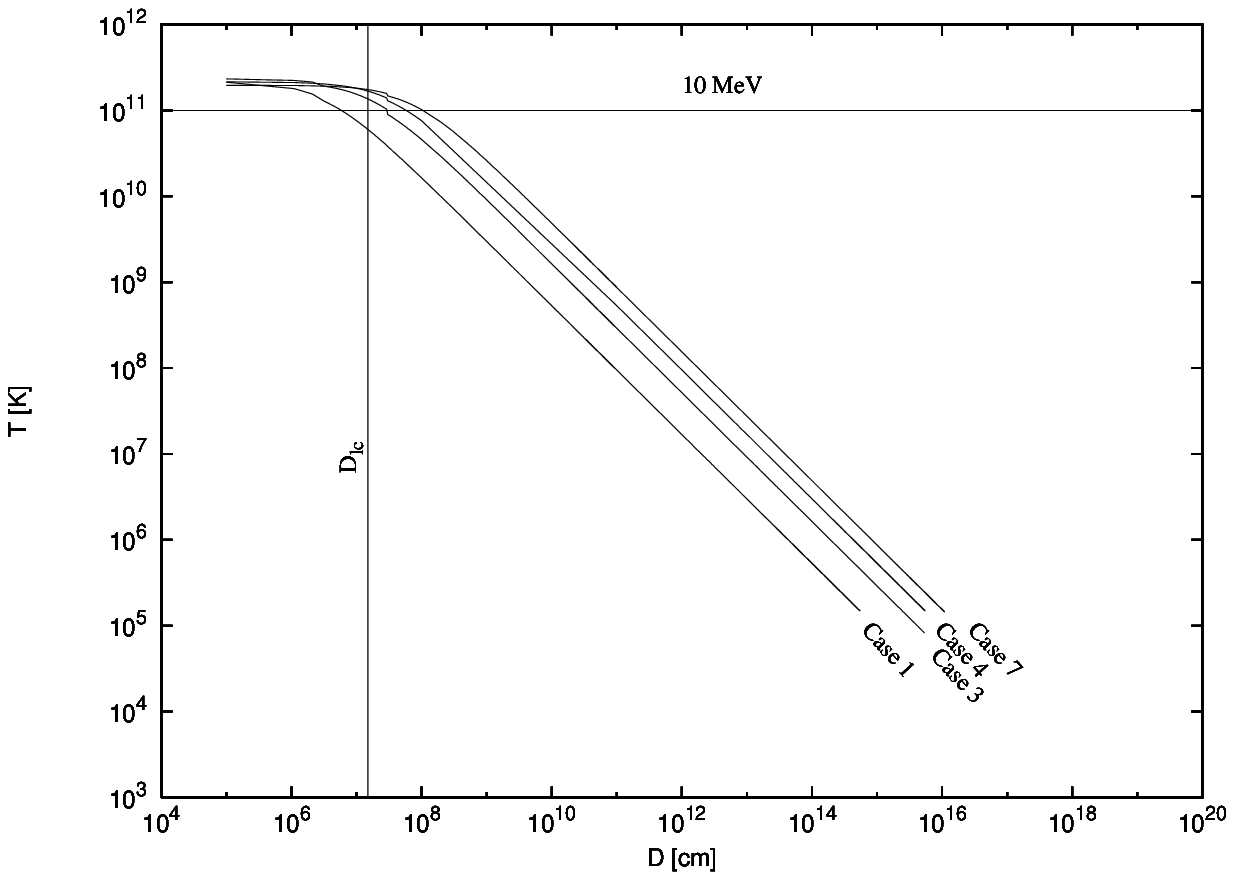}
\caption{Temperature vs distance from the origin for the CB, starting with hyperaccreting disk conditions.
 The forward integration is stopped when the CB becomes transparent to its enclosed radiation. The initial temperature for all cases is about $2.7\times10^{11}$ K. }
\label{forwardT}
\end{figure}

\begin{figure}
\includegraphics[scale=0.7]{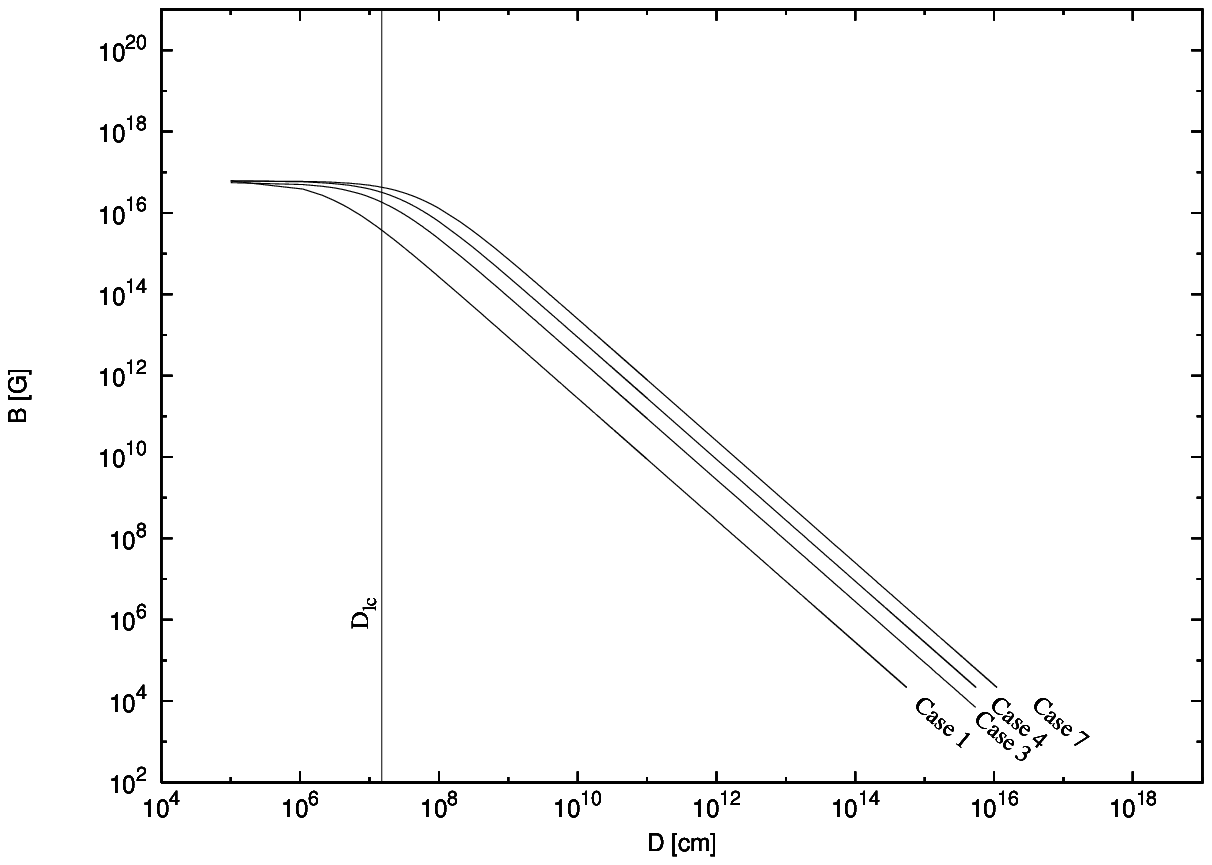}
\caption{Magnetic field vs distance from the origin for the CB, starting with hyperaccreting disk conditions.
 The forward integration is stopped when the CB becomes transparent to its enclosed radiation. The initial magnetic field for all cases is about $5\times10^{16}$ G. }
\label{Bforward}
\end{figure}

\begin{figure}
\includegraphics[scale=0.7]{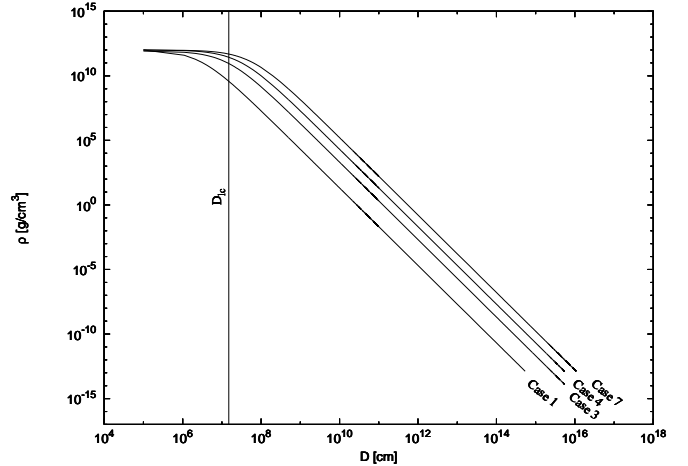}
\caption{Density vs distance from the origin for the CB, starting with hyperaccreting disk conditions.
 The forward integration is stopped when the CB becomes transparent to its enclosed radiation. The initial density for all cases is about $1\times10^{12} {\rm g/cm^3}$. }
\label{rhoforward}
\end{figure}
%------------------------------------------------------------------------------

For completeness and for self-consistency check, here we consider CBs with hyperaccreting disk conditions at the origin and perform a forward integration until the CBs reach the distances where they become transparent.

Assuming that the CB radius
 evolves as before,  $R=c/(\Gamma_{\rm CB}\sqrt{3})$,
then the density at a distance corresponding to the surface of the hyperaccreting disk ($D=10^5$ cm) will be too high. We will thus
 make a slight adjustment by rewriting the radius as $R=D/(\Gamma_{\rm CB}\sqrt{3})+x$, where $x$ is a number that ensures that the density
 at the origin does not exceed $\rho=10^{12} {\rm g/cm^3}$. Therefore, $x$ is found by solving the following equation:
\begin{equation}
10^{12}{\rm g/cm^3}=\frac{M_{\rm CB}}{\frac{4}{3}\pi \big(\frac{10^5 {\rm cm}}{\Gamma_{\rm CB}\sqrt{3}}+x\big)^3},
\end{equation}
which implies:
\begin{equation}
x[{\rm cm}]=\frac{3M_{\rm CB}^{1/3}(6/\pi)^{1/3}\Gamma_{\rm CB}-\sqrt{12}\times10^9}{60000\Gamma_{\rm CB}}.
\end{equation}
This also ensures the correct expansion velocity $v_{\rm exp}=c/\sqrt{3}$.
Table \ref{forwardx} shows the corresponding parameter values for
 $x$ and $\epsilon_{\rm m}$.
We should also note that in this case $\epsilon_{\rm m}$ will be chosen as to insure energy equipartition at the disk surface.
 The temperature thus found is used to calculate the neutrino emissivity, which is then added to the total energy in Eq.~\ref{degtot} to find the new temperature. As before, the neutrinos are released in small successive bursts mimicking a continuous
release of neutrinos. 

\begin{table}[h!]
\caption{The parameter $x$ used in the relation between $R_{\rm CB}$ and $D$, and $\epsilon_{\rm m}$ for the different CB cases when starting with disk conditions and integrating forward.}
\centering
\begin{tabular}{lll}
\hline
\noalign{\smallskip}
Case & $x$ [cm] & $\epsilon_{\rm m}$ \\
\noalign{\smallskip}
\hline \hline
\noalign{\smallskip}
1 & $15279.2$ & $0.045$ \\
2 & $33584.5$ & $0.045$ \\ 
3 & $73022.3$ & $0.045$ \\ 
4 & $15798.9$ & $0.045$ \\ 
5 & $34104.2$ & $0.045$ \\ 
7 & $15827.7$ & $0.045$ \\ 
\noalign{\smallskip}
\hline
\end{tabular}
\label{forwardx}
\end{table}

Figs.~\ref{forwardT}-\ref{rhoforward} shows the temperature, magnetic field and density as a function of distance.
All cases starts with $T\sim 2\times10^{11}$ K, $\rho=10^{12} {\rm g/cm^3}$ and $B=6\times 10^{16}$ G at $D=10^5$ cm. The neutrino effects can be seen as small jumps in the temperature curves, but in general the neutrinos do not change the overall picture a lot. The neutrino contribution were of the same order or smaller than the total energy, and as discussed before the temperature is not very sensitive to changes in the total energy.

The temperature at $D_{\rm trans}$ is of the same order as for the backward integration ($T=10^4$ K to $T=10^5$ K), and also close to the value given by Dado et al. (2002) of $T_{\rm trans}\simeq 4$ eV. The difference between the backward and forward integration at large distances is due to the different $\epsilon_{\rm m}$ parameter. For large distances, the $x$-parameter does not play any role.

To summarize, the results of the forward integration indicate that CBs formed within hyperaccretion disks could in principle provide the necessary conditions at $D_{\rm trans}$ to account for GRB features as claimed in Dado et al. (2002).


\begin{thebibliography}{}

\bibitem[1991]{balbus91} Balbus, S. A. \& Hawley, J. F. 1991, ApJ, 376, 214

\bibitem[dado]{dado02} Dado, S., Dar, A., \& De R{\'u}jula, A.
2002, A\&A,  388, 1079

\bibitem[Dar \& De R{\'u}jula 2004]{dar04}Dar, A. \& De R{\'u}jula, A. 2004, Phys. Rep. 405, 203

\bibitem[2005]{devilliers05}De Villiers, J. P., Staff, J., Ouyed, R. 2005, astro-ph/0502225

\bibitem[2002]{dimatteo02}Di Matteo, T., Perna, R., \& Narayan, R. 2002, ApJ, 579, 706

\bibitem[2001]{fendt01}Fendt, Ch. \& Memola, E. 2001, A\&A 365, 631

\bibitem[2004]{fendt04}Fendt, Ch. \& Ouyed, R. 2004, ApJ, 608, 378

\bibitem[1991]{hawley91}Hawley, J. F. \& Balbus, S. A. 1991, ApJ, 376, 223

\bibitem[1969]{michel69}Michel, F. C. 1969, ApJ, 158, 727

\bibitem[1995]{narayan95}Narayan, R. \& Yi, I. 1995, ApJ, 452, 710

\bibitem[1998]{narayan98}Narayan, R., Mahadevan, R., \& Quataert, E. 1998, in The Theory of Black Hole Accretion Discs, ed. M. A. Abramowicz, G. Bjornsson, \& J. E. Pringle (Cambridge: Cambridge Univ. Press),1998., p.148

\bibitem[2001]{narayan01}Narayan, R., Piran, T., \& Kumar, P. 2001, ApJ, 557, 949 

\bibitem[1996]{nishida96}Nishida, S. \& Eriguchi, Y. 1996, ApJ, 461, 320 

\bibitem[1997]{ouyed97}Ouyed, R., Pudritz, R. E., Stone, J. M. 1997, Nature, 385, 4090

\bibitem[1999]{piran99} Piran, T.\ 1999, Physics Reports, 314, 757

\bibitem[1999]{popham99}Popham, R., Woosley, S. \& Fryer, Ch. 1999, ApJ, 518, 356

\bibitem[1999]{rosswog99}Rosswog, S., Liebend\"orfer, M., Thielemann, F.-K., Davies, M.B., Benz, W.,
   \&  Piran, T. 1999, A\&A, 341, 499

\bibitem[1973]{shakura73}Shakura, N. I. \& Sunyaev, R. A. 1973, A\&A, 24, 337

\bibitem[1995]{spruit95}Spruit, H. C., Stehle, R. \& Papaloizou, J. C. B. 1995, MNRAS, 275, 1223

\bibitem[1999]{tagger92}Tagger, M., Pellat, R. \& Coroniti, F., 1992, ApJ 393, 708

\end{thebibliography}
\end{document}